\documentclass[twocolumn, aps, superscriptaddress, groupedaddress]{revtex4}  

\newcommand{\bra}[1]{\langle #1|}  
\newcommand{\ket}[1]{|#1\rangle}

\usepackage{mathtools}
\usepackage{lipsum}
  
\usepackage{amsmath,amsthm,amsfonts,amssymb}   
\usepackage{fontenc}
\usepackage[all]{xy}
\usepackage{graphicx,subfigure,multirow}
\usepackage{tikz}
\usetikzlibrary{shapes, arrows, shadows} 
\usepackage[latin1]{inputenc}
\usepackage[scaled]{helvet}
\usepackage[toc, page]{appendix}
\usepackage{graphicx}				
\usepackage{amssymb}

\newtheorem{result}{Result}

\begin{document} 

\title{Device-independent certification of non-classical joint measurements via causal models}   
\author{Ciar{\'a}n~M. Lee} 
\email{ciaran.lee@ucl.ac.uk}
\affiliation{Department of Physics and Astronomy, University College London, Gower Street, London WC1E 6BT, UK}

\begin{abstract}      
Quantum measurements are crucial for quantum technologies and give rise to some of the most classically counter-intuitive quantum phenomena. As such, the ability to certify the presence of genuinely non-classical joint measurements in a device-independent fashion is vital. However, previous work has either been non-device-independent, or has relied on post-selection\color{black}---the ability to discard all runs of an experiment in which a specific event did not occur\color{black}. In the case of entanglement, the post-selection approach applies an entangled measurement to independent states and post-selects the outcome, inducing non-classical correlations between the states that can be device-independently certified using a Bell inequality. That is, it certifies measurement non-classicality not by what it \emph{is}, but by what it \emph{does}. This paper remedies this discrepancy by providing a novel notion of what measurement non-classicality \emph{is}, which, in analogy with Bell's theorem, corresponds to measurement statistics being incompatible with an underlying classical causal model. It is shown that this provides a more fine-grained notion of non-classicality than post-selection, as it certifies the presence of non-classicality that cannot be revealed by examining post-selected outcomes alone. 
\end{abstract} 

\maketitle      

Quantum measurements are a key resource behind most quantum technologies \cite{sangouard2011quantum} and, moreover, they reveal some of the most startling non-classical features of quantum theory \cite{branciard2010characterizing, mazurek2016experimental}. Indeed, performing joint quantum measurements on composite systems is a key feature behind quantum teleportation, superdense coding, metrology, cryptography \cite{lee2018towards}, quantum repeaters \cite{sangouard2011quantum}, and quantum networks more generally. Hence the ability to certify the non-classical nature of quantum measurements is vitally important for the functioning of quantum technology and additionally, for understanding some of the fundamental differences between quantum and classical physics. Moreover, as the manufacturers of quantum measurement devices may not always be trusted, such certifications should be device-independent. That is, they should rely only on output measurement statistics rather than any intrinsic quantum properties, such as knowledge of the underlying Hilbert space dimension. 
 
Previous work on the certification of joint quantum measurements largely \footnote{Note that the work of Ref.~\cite{kaniewski2017self} gave a method to device-independently certify the presence of \emph{local} quantum measurements. However, this method was relational in the sense that it only certified how non-classical one local measurement was with respect to another. That is, it only certified how two local measurements relate to each other, but not what they are individually. In the current work, the case of individual joint quantum measurements (i.e a single measurement acting jointly on multipartite systems) is considered.} falls into two categories. The first uses witnesses to certify the presence of non-classical measurements \cite{vertesi2011certifying, bennet2014experimental, bowles2015testing}, but is manifestly not device-independent. The second is device-independent, but requires post-selection\color{black}---the ability to discard all runs of an experiment where a specific event did not occur---\color{black}to certify the presence of a quantum measurement \cite{rabelo2011device}. In the case of entanglement, such certification is accomplished by exploiting the fact that applying an entangled measurement to two initially independent entangled states and post-selecting the outcome, \color{black}that is considering only those runs in which a single fixed outcome occurred\color{black}, induces entanglement between the states, which can then be certified device-independently by violating a Bell inequality. This method hence detects quantum measurements through their action on states. That is, it certifies an entangled measurement through what is \emph{does}, not what it \emph{is}. This is in stark contrast with entangled states, whose non-classicality is easily certified through the violation of a Bell inequality. Such violation implies a denial of (at least one of) the assumptions underlying Bell's theorem. The modern treatment of which utilises the classical causal model framework to unify Bell's original assumptions \cite{wood2015lesson, chaves2015unifying, PhysRevX.8.021018}. Composite states are thus said to be non-classical if the correlations generated by locally measuring each composite system are inconsistent with an underlying classical causal model.     
  
This paper remedies the discrepancy between the treatment of non-classicality in quantum states and measurements. In analogy with Bell's theorem, a joint quantum measurement is said to be non-classical if the correlations generated by performing it on local preparations on each composite system are inconsistent with an underlying classical causal model. In the following section this classical causal model is introduced and a non-linear inequality on any distribution generated by it is derived. Violation of this inequality entails that the observed correlations are in conflict with the classical causal model. As the inequality depends only on observed output statistics, it is manifestly device-independent. Additionally, it will be demonstrated that this inequality provides a finer-grained notion of joint measurement non-classicality for general quantum measurements than the post-selection approach of Ref.~\cite{rabelo2011device}, discussed above, as it certifies the presence of non-classicality that cannot be revealed by examining correlations arising from post-selecting the outcomes of measurement alone.
  
\begin{figure}[t]
\includegraphics[scale=.42]{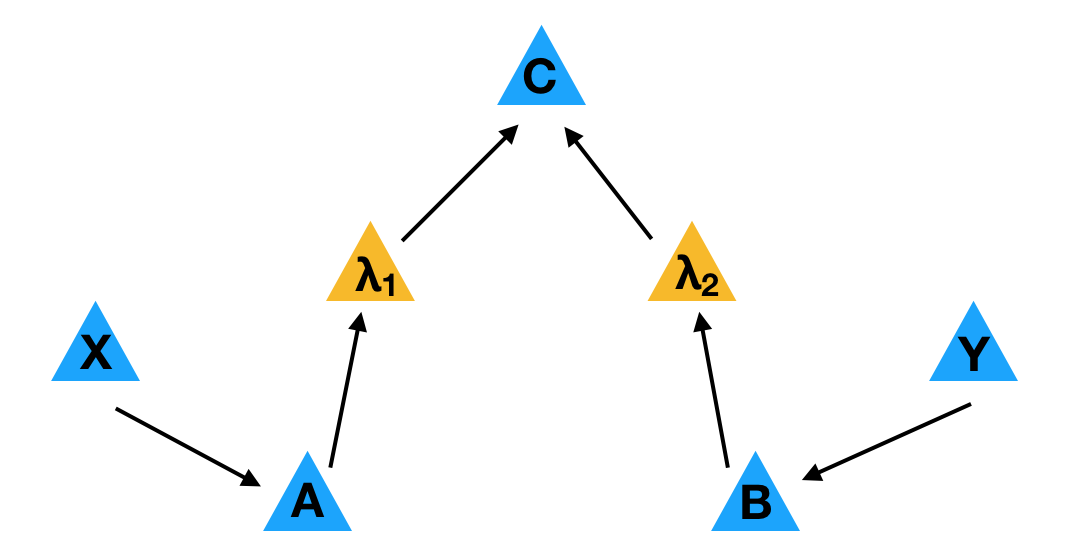}  
\caption{DAG representing classical causal model. \color{black} This causal structure is similar to the bilocality structure introduced in Ref.~\cite{branciard2012bilocal}, with a few key differences. Firstly, Alice and Bob have preparation devices, rather than measurement devices. Moreover, in the structure considered here, there is an arrow from preparation outcome to the hidden variable rather than the other way around---as is the case in the bilocality set-up. Hence, here, the hidden variables can a priori depend on the choice of preparation. That is, it does \emph{not} follow from the above DAG that $P(\lambda_1|x)=P(\lambda_1)$. \color{black}} 
\label{one}
\end{figure}

\emph{Certifying non-classical joint measurements.---} Recently, tools and techniques from the classical causal models framework have begun to see myriad applications in quantum information \cite{wood2015lesson, allen2016quantum, chaves2015device, chaves2015unifying, chaves2016causal, brask2017bell, PhysRevX.8.021018, lee2015causal}\footnote{For connections between related notions of causality and quantum information, see \cite{lee2016deriving, lee2017no, lee2016bounds, lee2016information, barrett2017computational, barnum2018oracles, barnum2017ruling, lee2015computation, lee2016generalised}}. In this framework, the inputs and outputs of agents measurement and preparation devices are represented by nodes in directed acyclic graphs (DAGs), with the arrows denoting the causal relationship between nodes. The structure of each DAG encodes conditional independence relations \footnote{Here the \emph{faithfulness} condition is being assumed, see \cite{Pearl-09,wood2015lesson} for a discussion.} among the nodes. For instance, the no-signalling conditions $P(A|X,Y)=P(A|X) \text{, and } P(B|X,Y)=P(B|Y)$, follow directly \cite{wood2015lesson} from the structure of the DAG from in Fig.\ref{one}. Indeed, the sructure of the DAG specifies all the conditional independences between the nodes \cite{Pearl-09, henson2014theory}. In short, every relation between the inputs and outputs of the different agents are specified by the DAG.  

Consider three agents Alice, Bob, and Charlie. Alice and Bob both have devices which prepare a quantum state from some ensemble of states, given a choice between different ensembles. Charlie has a measurement device which jointly measures the states prepared by Alice and Bob. The actions of these devices are represented in a black-box manner. Alice and Bob's devices have a classical input $x,y$ (the choice of different ensemble) respectively, and a classical output $a, b$ (the state prepared from the chosen ensemble) respectively. Here it is assumed that $a,b,x,y \in \{0,1\}$. Charlie has no classical input, as his device only performs a single measurement, but has a classical output $C$ indexing the possible measurement outcomes. It is assumed in this section that $C$ takes four values and hence is indexed by two bits, $C=c_0c_1 \in \{00, 01, 10, 11\}$. Preparing and measuring states in this manner gives rise to a conditional probability distribution $P(a, b, c_0c_1 | x,y)$.
 
In analogy with Bell's theorem, a classical causal model for $P(a, b, c_0c_1 | x,y)$ is described by the DAG in Fig.~\ref{one}, where $\lambda_1, \lambda_2$ are unobserved, independent random variables. If the correlations generated by performing Charlie's measurement on Alice and Bob's preparations are consistent with the DAG in Fig.~\ref{one}, then they are said to be classical. That is, they are mediated by the hidden random variables $\lambda_1, \lambda_2$. One might wonder why there are two hidden variables, rather than one. This is due to the independence of Alice and Bob's devices: $P(a,b|x,y)=\sum_{c_0c_1}P(a,b,c_0c_1|x,y)=P(a|x)P(b|y)$. If the correlations between Alice, Bob, and Charlie were mediated by a single hidden variable, then Alice and Bob's marginal distribution would not be independent. A bound on the possible classically generated correlated is now presented. 

\begin{result} 
A distribution $P(a, b, c_0c_1 | x,y)$ generated by the DAG of Fig.~\ref{one} satisfies:
$$ \sqrt{|M|} + \sqrt{|N|} \leq 1, $$ 
$$ 
\begin{aligned}
\text{where }\qquad M&=\frac{1}{4}\sum_{xy} \langle A_xB_yC^0\rangle, \\ 
\text{and } \quad\qquad N&=\frac{1}{4}\sum_{xy} (-1)^{x+y} \langle A_xB_yC^1\rangle, \\
\text{and } \langle A_xB_yC^i \rangle &= \sum_{ab{c_0} c_1} (-1)^{a+b+c_i} P(a,b,c_0c_1|x,y). 
\end{aligned}
$$
\end{result} 

Note that, in contrast to standard Bell inequalities, the inequality presented above is non-linear in the joint distribution $P(a,b,c_0c_1|x,y)$. This is due to the independence of Alice and Bob's preparations. 

The proof of Result 1 is similar to the derivation of the bilocality inequality from Ref.~\cite{branciard2012bilocal}, with a few key differences. First, in the case considered here, the hidden variables can a priori depend on the choice of preparation. That is, it does not follow from the DAG of Fig.~\ref{one} that $P(\lambda_1|x)=P(\lambda_1)$. Lastly, Alice and Bob have preparation devices, rather than measurement devices.

\emph{Proof.}
Given the structure of the DAG from Fig~\ref{one}, it follows that $P(a,b,c_0c_1|x,y)$ decomposes as
$$ \iint d\lambda_1 d\lambda_2 P(a|x)P(b|y)P(c_0c_1|\lambda_1 \lambda_1)P(\lambda_1|x)P(\lambda_2|y). $$
Define $\langle A_x\rangle=\sum_a(-1)^aP(a|x)$, $\langle B_y\rangle=\sum_b(-1)^aP(b|y)$, and $\langle C^i\rangle_{\lambda_1 \lambda_2}=\sum_{c_0 c_1}(-1)^{c_i}P(c_0c_1|\lambda_1 \lambda_2).$ It follows from the above decomposition that one can write $\langle A_xB_yC^i\rangle$ as
$$\iint d\lambda_1 d\lambda_2 \langle A_x \rangle \langle B_y \rangle \langle C^i\rangle_{\lambda_1 \lambda_2} P(\lambda_1|x)P(\lambda_2|y).  $$
This, together with $|\langle C^i\rangle_{\lambda_1 \lambda_2}| \leq 1 $, implies

$$ 
\begin{aligned}
 |M| \leq & \left( \int d\lambda_1 \frac{| \langle A_0 \rangle P(\lambda_1|0) + \langle A_1 \rangle P(\lambda_1|1) |}{2} \right)\cdot  \\
&\quad\quad\quad \cdot \left( \int d\lambda_2 \frac{| \langle B_0 \rangle P(\lambda_2|0) + \langle B_1 \rangle P(\lambda_2|1) |}{2} \right).
 \end{aligned}$$
 
One can similarly bound $|N|$. The key difference being that the $+$'s in the above bound are replaced with $-$'s due to the occurrence of the $(-1)^{x+y}$ term in $N$. For real  $z, w, z', w' \geq 0$, it was proved in Ref.~\cite{branciard2012bilocal} that the inequality $\sqrt{zw} + \sqrt{z'w'} \leq \sqrt{z+z'}\sqrt{w+w'}$ holds. Hence 

\begin{widetext}
$$
\begin{aligned}
\sqrt{|M|} + \sqrt{|N|} \leq &  \sqrt{ \int d\lambda_1 \left( \frac{| \langle A_0 \rangle P(\lambda_1|0) + \langle A_1 \rangle P(\lambda_1|1) |}{2} + \frac{| \langle A_0 \rangle P(\lambda_1|0) - \langle A_1 \rangle P(\lambda_1|1) |}{2} \right)} \cdot \\ 
& \quad \quad \cdot  \sqrt{ \int d\lambda_2 \left( \frac{| \langle B_0 \rangle P(\lambda_2|0) + \langle B_1 \rangle P(\lambda_2|1) |}{2} + \frac{| \langle B_0 \rangle P(\lambda_2|0) - \langle B_1 \rangle P(\lambda_2|1) |}{2} \right)} \\
\leq & \sqrt{ \int d\lambda_1  \max \Big( |\langle A_0 \rangle P(\lambda_1|0)|, \langle A_1 \rangle P(\lambda_1|1) \Big) } 
 \cdot  \sqrt{ \int d\lambda_2 \max \Big( |\langle B_0 \rangle P(\lambda_2|0)|, \langle B_1 \rangle P(\lambda_2|1) \Big) } \leq 1 \qquad \square
\end{aligned}
$$
\end{widetext} 

The bound from Result 1 can be classically saturated. To see this, consider the following. Let $x,y$ be independent, uniformly distributed random bits. Let Alice's (Bob's) device output $a=x\oplus 1$ ($b=y \oplus 1$) with probability one, and let $\lambda_1$ ($\lambda_2$) equal $a\oplus 1$ ($b\oplus 1$) with probability one. Let Charlie have two independent and identically distributed random bits $\mu_0$ and $\mu_1$. When both $\mu_0$ and $\mu_1$ equal zero, Charlie's device outputs $(c_0,c_1)=(\lambda_1 \oplus \lambda_2, \nu)$ with probability one, where $\nu$ is another random bit. When $\mu_0$ and $\mu_1$ equal one, Charlie's device outputs $(c_0,c_1)=(\nu, \lambda_1 \oplus \lambda_2)$ with probability one. When $\mu_0 \neq \mu_1$ Charlie's device outputs $(c_0,c_1)=(\lambda_1, \lambda_2,)$ with probability one. When $\mu_0=\mu_1=0$ it follows by a straightforward calculation that $M=1$ and $N=0$, and when $\mu_0=\mu_1=1$, $M=0$ and $N=1$. In all remaining cases $M=N=0$. As the probability that $\mu_0=\mu_1=0$ is $r^2$ and the probability that $\mu_0=\mu_1=1$ is $(1-r)^2$, where $r=P(\mu_0=0)=P(\mu_1=0)$, all points $(M,N)=(r^2, (1-r)^2)$ can be achieved. The boundary $\sqrt{|M|} + \sqrt{|N|} =1$ is thus classically saturated.  

\emph{Quantum violation.---} \color{black} Recall that Alice and Bob's devices prepare a single state from an ensemble of two states, given a choice between two possible ensembles. For the quantum violation of the bound from Result 1, the preparation device used by Alice and Bob is the same. The functioning of this device will now be specified. For $x=0$ ($y=0$, respectively) one of the two states from the basis
$$
\{ \cos\left(\frac{\pi}{8}\right) \ket{0} + \sin\left(\frac{\pi}{8}\right) \ket{1}, \cos\left(\frac{\pi}{8}\right) \ket{0} - \sin\left(\frac{\pi}{8}\right) \ket{1} \}
$$
is prepared, with the value of $a$ ($b$)---the preparation outcome---denoting whether the first ($`0'$th) or second ($`1'$st) state is prepared. Each state is equally likely. For $x=1$ ($y=1$) one of the two states from the basis
$$
\{ \cos\left(\frac{3\pi}{8}\right) \ket{0} + \sin\left(\frac{3\pi}{8}\right) \ket{1}, \cos\left(\frac{3\pi}{8}\right) \ket{0} - \sin\left(\frac{3\pi}{8}\right) \ket{1} \}
$$
is prepared. Again, the value of $a$ ($b$) denotes whether the first ($`0'$th) or second ($`1'$st) state is prepared. For instance, if Alice chooses preparation $x=1$ and observes outcome $a=0$, then the state 
$$ \cos\left(\frac{3\pi}{8}\right) \ket{0} + \sin\left(\frac{3\pi}{8}\right) \ket{1} $$
has been prepared by her device. It is clear that the states received by Charlie depend on the choice of preparation.

\begin{figure}[b]
\includegraphics[scale=.37]{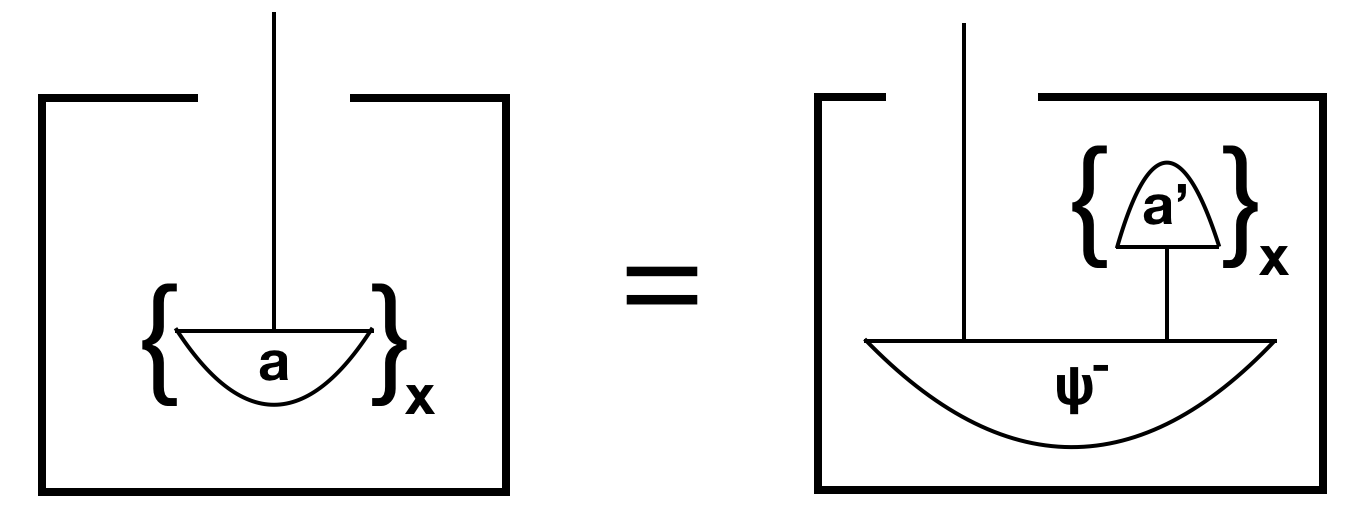}   
\caption{\color{black} Physical realisation of Alice and Bob's preparation devices. The box corresponds to the preparation device, and the vertical line emerging from it is the system Alice (Bob) sends Charlie. Here, on the left hand side of the diagram, $x$ denotes the different possible ensembles and $a$ denotes the specific state prepared from each ensemble. As Alice and Bob control both system and ancilla on the right hand side diagram, they know the measurement outcome $a'$. Hence, they know the exact state prepared on their original system. More specifically, the choice of measurement on their ancilla, denoted by $x$, specifies the two different ensembles the original system can be steered to. Moreover, the specific measurement outcome $a'$ prepares a fixed state $a$ from the ensemble associated with that choice of measurement.\color{black}} 
\label{preparation} 
\end{figure} 
 
A simple quantum realisation of Alice's (Bob's) preparation device is to prepare a maximally entangled $\ket{\psi^-}$ state between Alice's (Bob's) system and an ancilla, and perform a measurement on the ancilla to prepare a state on Alice's (Bob's) system. Given two distinct measurements that can be performed on the ancilla, there are two distinct ensembles of states to which Alice's (Bob's) system can be steered. The specific measurement outcome prepares a fixed state from the ensemble associated with that measurement. Note that as Alice has control of both her original system and the ancilla, she knows the measurement outcome on the ancilla system and hence what state is prepared on her system. There is hence a mathematical correspondence between the choice of measurement on the ancilla and the choice of preparation on the original system. In effect, when a single agent holds both systems, measuring one half of an entangled state is mathematically equivalent to preparing a state on the other system. This is schematically depicted in Fig.~\ref{preparation}, see the caption for further details. 

Now, to achieve the specific state preparations described at the start of this section, Alice (Bob) performs either $(\sigma_Z+\sigma_X)/\sqrt{2}$ (for $x=y=0$) or $(\sigma_Z-\sigma_X)/\sqrt{2}$ (for $x=y=1$) on her (his) ancilla. This provides a concrete physical implementation of the  preparation devices held by Alice and Bob, described earlier in this section. \color{black}

Finally, Charlie performs the `noisy' Bell state measurement $\{E_{c_0c_1}\}$ on his system, where $E_{c_0c_1}= p \ket{\psi_{c_0c_1}}\bra{\psi_{c_0c_1}} +(1-p)\mathbb{I}/4$, and $\{\ket{\psi_{00}}\bra{\psi_{00}}, \ket{\psi_{01}}\bra{\psi_{01}}, \ket{\psi_{10}}\bra{\psi_{10}},\ket{\psi_{11}}\bra{\psi_{11}}\}$ is the Bell state measurement. As $E_{c_0c_1}\geq 0, \forall c_0c_1$, and $\sum_{c_0c_1} E_{c_0c_1} =\mathbb{I}$, this is a valid measurement. 

The correlations generated by the above preparation and measurement procedure are the same as those considered in Section III A of Ref.~\cite{branciard2012bilocal}, namely:
$$
\begin{aligned}
&P(a,b,c_0c_1|x,y) = \\
& \quad\quad \frac{1}{16} \left( 1 + p(-1)^{a+b} \Big\{\frac{(-1)^{c_0} + (-1)^{x+y+c_1}}{2} \Big\} \right). 
\end{aligned}
$$
From this one obtains $ \sqrt{|M|} + \sqrt{|N|} = \sqrt{2p},$ providing a quantum violation for $p>1/2$.

\emph{Post-selection.---} Ref.~\cite{rabelo2011device} demonstrated that the presence of an entangled measurement can be certified in an device-independent fashion using post-selection. This was achieved by exploiting the fact that performing an entangled measurement on two initially independent entangled states and post-selecting the outcome\color{black}---that is considering only those runs of the experiment in which a specific fixed outcome occurs---\color{black}induces entanglement between the states, which can then be certified device-independently by violating a \color{black}Bell inequality \footnote{\color{black}Note that post-selection here does not refer to finite sampling effects. It is hence not related to the fair sampling loophole in Bell experiments, which concerns practical limitations on the efficiencies of measurement devices\color{black}}\color{black}. This method detects entangled measurements through their action on states by showing that for each fixed measurement outcome the induced correlations are non-classical. In the current work a novel method has been introduced which certifies general measurement non-classicality not through what it \emph{does}, but what it \emph{is}. These two approaches coincide for entangled measurements \cite{rabelo2011device}, but do they coincide for general non-classical measurements? That is, if a measurement is non-classical in the sense that it violates the bound from Result 1, are the correlations induced between Alice and Bob's devices on post-selection of Charlie's outcome always non-classical? It will now be shown that, surprisingly, the existence of a separate classical model for each post-selected measurement outcome does not imply the measurement is classical in the sense of Fig.~\ref{one}. 
 
Note that given the realisations, introduced in the previous section, of Alice and Bob's preparation devices involving steering using projective measurements on an ancilla, it follows that non-classical correlations between Alice and Bob's preparation devices are equivalent to non-classical correlations between projective measurements performed on their ancillas. 

Now, consider the following. Allow Charlie to perform a noisy Bell state measurement with noise parameter $p$ and post-select on an arbitrary fixed outcome. If Alice and Bob each have their own Bell state, then Charlie's joint measurement on two of their systems induces a noisy Bell state---with the same noise parameter $p$---between Alice and Bob's ancilla. For instance, if Charlie post-selects outcome $E_{00}= p \ket{\psi^-}\bra{\psi^-} +(1-p)\mathbb{I}/4$, then Alice and Bob's ancilla will be in the $p \ket{\psi^-}\bra{\psi^-} +(1-p)\mathbb{I}/4$ state. Hence, classically simulating Charlie's joint noisy Bell measurement on Alice and Bob's preparations is equivalent to classically simulating local projective measurements on Alice and Bob's ancilla's in the induced noisy Bell state. \color{black} As shown in \cite{hirsch2017better}, such correlations can be classically simulated for $p<0.68$\color{black}.  But, as shown in the previous section, the non-post-selected measurement is non-classical as long as $p>1/2$. To summarise, the following has been shown:

\begin{result}
The existence of a separate classical model for each joint measurement outcome---adhering to the constraints imposed by Fig.~\ref{one}---does not imply the joint measurement is classical in the sense of Fig.~\ref{one} and Result 1. 
\end{result}

An intuitive explanation of this result could be that, as Charlie's measurements outcomes can overlap on certain states, classical models for each individual measurement outcome cannot always be combined consistently.

\begin{figure}[t]
\includegraphics[scale=.42]{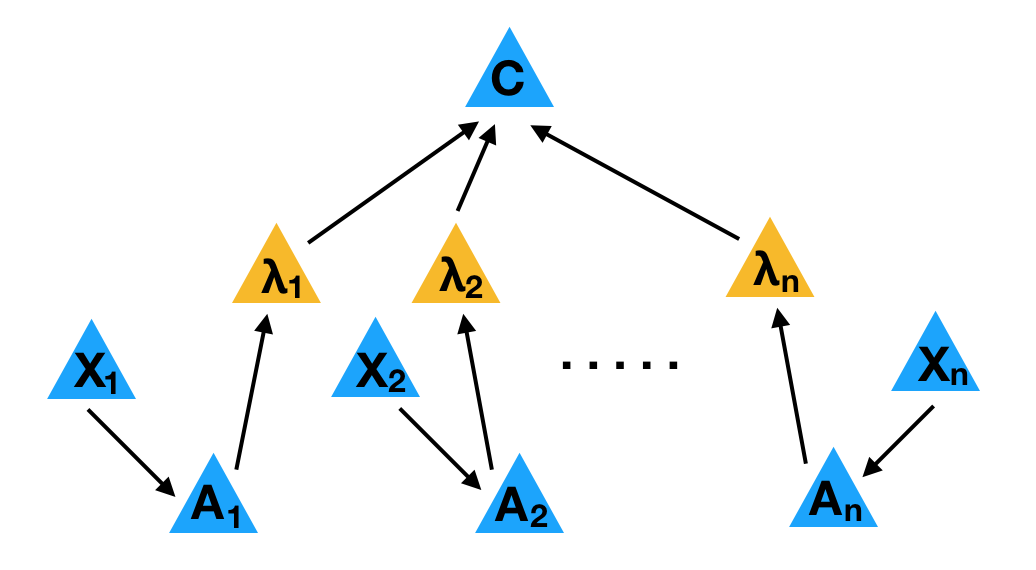}
\caption{DAG for composite joint measurement} 
\label{multi}
\end{figure}

\emph{Generalisation to $n$ systems and $2^k$ outcomes.---} The inequality from Result 1 will now be generalised to allow for $n$ systems, $k$ choices for the each preparation device---each of which have two possible outcomes---and $2^k$ possible outcomes for Charlie's joint measurement, indexed using $k$ bits $c_0\cdots c_{k-1}$. Result 1 corresponds to the $n=k=2$ case. As before, the classical causal model is depicted in Fig.~\ref{multi}.

\begin{result}
A distribution $$P(a_1, \dots, a_{n}c_0\cdots c_{k-1}|x_1,\dots, x_{n}),$$ with $a_i, c_j \in\{0,1\}$ and $x_i\in\{0,\dots, k-1\}$, generated by the DAG of Fig.~\ref{multi} satisfies the following inequality:
$$
\mathcal{S}:= \sum_{i=0}^{k-1} {|I_i|}^{1/n} \leq k-1,
$$
$\text{where, } I_i = \frac{1}{2^n}\sum_{x_1,\dots,x_n=i}^{i+1} \langle A^1_{x_1} \cdots A^n_{x_n} C^i \rangle$, for $i$ ranging from  $0 \text{ to } k-1$, with $A^i_{k}=-A^i_0$ and
$\langle A^1_{x_1} \cdots A^n_{x_n} B_y \rangle = \sum (-1)^{b_i+\sum_{j=1}^n a_j} P(a_1, \dots, a_{n}c_0\cdots c_{k-1}|x_1,\dots, x_{n})$.
\end{result}

\emph{Proof.}
Given the decomposition of the distribution over the agents preparations and Charlie's measurement, $$P(a_1, \dots, a_{n}c_0\cdots c_{k-1}|x_1,\dots, x_{n}),$$ implied by the structure of Fig~\ref{multi}, it follows that
$$
|I_i| \leq \prod_{j=1}^n \Big( \frac{1}{2} \int  \Big| \sum_{x_j=1}^n \langle A^j_{x_j} \rangle  p(\lambda_j|x_j) \Big| d\lambda_j \Big),
$$
where $ \langle A^j_{x_j} \rangle = \sum_{a_j} (-1)^{a_j} P(a_j|x_j)$.

It was shown in Ref.~\cite{tavakoli2014nonlocal} that, for $c_i^k\in\mathbb{R}_+$ and $m,n\in\mathbb{N}$, the following holds:
$$
\sum_{k=1}^m\Big( \prod_{i=1}^n c_i^k \Big)^{1/n} \leq \prod_{i=1}^{i+1} \left( c_i^1+c_i^2+\cdots + x_i^m \right)^{1/n}.
$$
Applying this result to $\mathcal{S}=\sum_{i=0}^{k-1} {|I_i|}^{1/n}$ yields
$$
\mathcal{S} \leq \Big[ \prod_{j=1}^n \frac{1}{2} \int  \Big( \Big| \langle A^j_0 \rangle p(\lambda_j|0) + \langle A^j_1 \rangle p(\lambda_j|1) \Big| + \cdots  $$

$$\cdots + \Big| \langle A^j_{k-1} \rangle p(\lambda_j|k-1) - \langle A^j_0 \rangle p(\lambda_j|0) \Big|  \Big) d\lambda_j \Big]^{1/n}.
$$
The following upper bound holds:
$$
\begin{aligned}
\frac{1}{2} &\Big(\Big| \langle A^j_0 \rangle p(\lambda_j|0) + \langle A^j_1 \rangle p(\lambda_j|1) \Big| +  \\
&\cdots + \Big| \langle A^j_{k-1} \rangle p(\lambda_j|k-1) - \langle A^j_0 \rangle p(\lambda_j|0) \Big| \Big) \leq k-1.
\end{aligned}
$$
Hence, one has
$$
\mathcal{S} \leq \Big( \prod_{j=1}^n \int \left( k-1 \right)^n d\lambda_j \Big)^{1/n}  = k-1  \qquad\qquad \square$$

\emph{Conclusion.---} 
This paper has introduced a novel notion of non-classicality for joint quantum measurements. This notion took its cue from Bell's theorem and the device-independent certification of entangled quantum states by stipulating a joint quantum measurement to be non-classical if the correlations generated by performing it on local preparations are inconsistent with an underlying classical causal model. A non-linear inequality was then derived as a witness for this inconsistency: a violation entails non-classicality. This inequality bounded the classically generated correlations achievable with this causal model. In future work it would be interesting to investigate the corresponding bounds for LOCC, unentangled, and entangled measurements, as was done in the semi-device independent case by Ref.'s~\cite{vertesi2011certifying, bennet2014experimental}. 

Moreover, this approach was shown to provide a more fine-grained notion of non-classicality than the post-selection method of Ref.~\cite{rabelo2011device}. That is, there exists quantum joint measurements which admit a classical hidden variable model for each post-selected measurement outcome, but which are nevertheless non-classical and violate the inequality from Result 1. It would be interesting to determine if a quantum protocol exhibiting an information-theoretic advantage due to this discrepancy existed. That is, can an agent with access to the entire collection of correlations generated by a quantum joint measurement gain an advantage over an agent who only has access to a post-selected subset of those correlations? 

In future work, connections between the notion of non-classicality introduced here and that of contexuality discussed in Ref.~\citep{PhysRevX.8.021018} will be explored. \color{black}Moreover, possible extensions to other experimental set-ups involving joint quantum measurements---such as the `triangle scenario' studied in Ref.'s~\cite{fraser2018causal,gisin2018entanglement}---will also be explored. \color{black}

\color{black} There has recently been a surge of interest in self-testing entangled measurements \cite{renou2018self,bancal2018noise}. While these methods provide robust methods to certify the presence of entangled measurements, they do not provide a clear definition of non-classicality for general joint quantum measurements---as Bell's theorem does for entangled states. The current work remedied this situation by providing a clean notion of when a joint measurement should be said to be non-classical. Future work will look at the connections between these different approaches. 

Finally, it is hoped that the current work will lead to further fruitful applications of the causal model framework to research in quantum information. \color{black}

\emph{Acknowledgements.---} CML thanks Dan Browne for helpful discussions and J. J. Barry for encouragement. This work was supported by the EPSRC through the UCL Doctoral Prize Fellowship. 

\emph{Competing interests.---} The author declares that there are no competing interests.


\bibliographystyle{ieeetr}  
\bibliography{library} 
\end{document}